# Correlations in Single-Photon Experiments

**Sándor Varró**
Research Institute for Solid State Physics and Optics
H-1525 Budapest, PO Box 49, Hungary
E-mail: varro@sunserv.kfki.hu

**Abstract.** Correlations of detection events in two photodetectors placed at the opposite sides of a beam splitter are studied in the frame of classical probability theory. It is assumed that there is always only one photon present in the measuring apparatus during one elementary experiment (one measurement act). Due to the conservation of energy, there is always a strict anticorrelation in one elementary experiment, because the photon cannot excite both of the detectors at the same time. It is explicitely shown in several examples that the "bunching" and "anti-bunching" of the counts in serieses of elementary single-photon experiments is governed by the statistical properties of grouping the sequences of the elementary measurements.

**1. Introduction**

Since Einstein (1905) introduced the concept of "light quanta" (nowadays they are called "photons") there has been a wide-spreading investigation carried out to check whether one photon can interfere with itself, or, perhaps, it can be split. He wrote in the introduction of his path-breaking paper that "*According to the assumption to be kept in eye here, by spreading from a point in the outgoing light rays the energy is not distributed continuously to larger and larger spatial regions, but these rays consist of a finite number of energy quanta localized in spatial points which move without falling apart, and they can be absorbed or created only as a whole.*" This extreme particle picture for the photon (as a point-like singularity), deduced from the thermodynamical study of black-body radiation in the *Wien limit*, was refined in a later paper by Einstein (1909a) where he derived from the *exact Planck law* his famous fluctuation formula, which contains both particle-like and wave-like fluctuations. This was the first mathematically correct formula on the *wave-particle duality*. Einstein (1909b) wrote in the summary of his talk delivered at the famous Salzburg Meeting that "*Nevertheless, for the time being, the most natural notion seems to me, that the appearance of the electromagnetic fields of light would also be attached to singular points, like in the case of electrostatic fields according to the electron theory. It is not excluded that in such a theory the energy of the electromagnetic field could be viewed as localized in these singularities, like in the old action-at-a-distance theory. I think of such singular points surrounded by force fields, which, in essence are of a character of plane waves, whose amplitudes decrease by the distance from the singular points. If there are many such singularities in a region, then their force fields will be on top of each other, and this assembly will form an undulatory force field, which, perhaps could hardly be distinguished from an undulatory field in the sense of the present theory of light. Needless to say, such a picture is of no value until it leads to an exact theory. With the help of it I merely wanted to illustrate in short that each of the structural properties (the undulatory structure and the quantal structure) which both show up according to Planck's formula, should not be viewed as incompatible to each other.*" The details of this early developments can be found, for instance, in our recent work Varró (2006). The quantization of the radiation field in modern sense was first presented by Dirac (1927), which serves nowadays as a basic element in all texbooks on quantum electrodynamics and quantum optics (see e.g. Loudon 1973, Scully and Zubairy 1997 and Schleich 2001). According to the modern concept *"Each photon then interferes only with itself. Interference between two different photons never occurs."*, as Dirac (1947) stated in his famous book on quantum mechanics. In our view this statement is based on that in the modern theory of the quantized electromagnetic radiation field the spatial distribution of a quantized mode is determined according to classical electrodynamics and the usual boundary condition. Thus, regardless of how large or small the excitation degree (occupation number, which is 1 for single-photon experiments) of a particular mode is, the interference pattern has already been "encoded" in the true mode function, which takes into account the boundary conditions determined by the *whole* measuring apparatus. On the evolution of the basic quantum concepts and the modern photon concept see e.g. Pipkin (1978) and the comparative study by Kidd et al (1989).

Taylor (1909) was the first who studied experimentally whether a very low-intensity light beam could produce *interference*. Later Dempster and Batho (1927) investigated the same problem by using an echelon grating, and they concluded that the interference pattern survives even at very low intensities, like in the classical theory of Maxwell fields (see also Landé and Gerlach 1926). The interference phenomena of light at very low intensities have been analysed with the help of a more sophisticated experimental apparatus (by using photomultipliers) by Jánossy and Náray (1957) with the same conclusion (see also Jánossy 1973, and an earlier comparative study by Reynolds et al. 1969 on the interference effects produced by single photons). In this context it is interesting to note that Franson and Potocki (1988) observed single-photon interference (by using a single-atom source) over 45 meters by using a Jamin interferometer.

Another branch of investigations concerns the "splitting of the photon", or in other words, the distribution of the photon energy *hν* over several active charges, e.g. oscillators. To our knowledge, this question was first investigated experimentally by Gans and Miguez (1917) where they studied the refractivity of a glas lens at very low incoming light intensities. According to preliminary expectations, as they wrote in the introduction of their paper, one would expect that, since one photon cannot excite the bulk of the lens, it would simply pass through, without any changes. In contrast, according to their experimental results, the refractivity was completely normal, even when the intensity was so low that the average energy of the oscillators (representing the active radiators in the glas) was about only of order of $10^{-30}$ of one light quantum energy. Later Ádám et al. (1955) studied the intensity *correlation* between two coherent light beams produced by splitting the light beam by means of a semi-transparent mirror. The important new technical element in this experiment was the use of photomultipliers as detectors. Their conclusion was that "photons falling on the semi-transparent mirror are not split – but at random pass on in the one or the other components of the beam." From their evaluation of their experimental results it was concluded that if there were photons in the beam which were split, their relative rate could not have exceeded 0.6% of the total incoming intensity. Brannen and Ferguson (1956) performed soon a similar experiment, with the difference that they introduced a time delay in one arm of the measuring apparatus in order to reduce the number of accidental coincidences. They concluded that there is no correlation (less than 0.01%) between photons in coherent light rays. Later Clauser (1974) reinvestigated the earlier photon correlation experiments, and concluded, that due to technical reasons, they were not decisive to distiguish between the validity of classical and quantum theory. He measured various coincidence rates between four photomultipliers viewing cascade photons on opposite sides of dielectric beam-splitters. His results contradict to the prediction of any classical or semiclassical theory. The possible splitting of a single photon has been analysed recently by Ghose et al. (1991) and Mizobuchi and Ohtaké (1992) by using a double prism instead of a dielectric beam-splitter. Their results are in accord with the modern quantum theory of radiation. Going back to the early days of photon correlation experiments, we have to mention the work by Hanbury Brown and Twiss (1956) in which they developed a new type of radio interferometer based on the measurement of the correlation of the intensity fluctuation at the two detectors, which has been used to measure the angular diameter of radio stars. Later they built up an analogon of this device working at optical wavelengths and measured the correlations in a laboratory experiment. They measured positive correlation in this experiment, which they interpreted on the basis of semiclassical radiation theory (Hanbury Brown and Twiss 1957, see also Jánossy 1957). A similar experiment was performed by Rebka an Pound (1957) on time-correlated photons. In a theoretical study of intensity fluctuations in stationary optical field, Wolf (1957) found also positive correlation on the basis of analysing coupled classical Gaussian random processes, thus served an interpretation of the Hanbury Brown and Twiss effect (photon bunching). Later Arecchi et al. (1966) performed the cleanest experiments at that time to compare the time distribution of photons from coherent and Gaussian sources. In case of a coherent source the intensity-intensity correlation is insensitive from the time delay (its normalized value is unity). On the other hand for a Gaussian source the normalized intensity-intensity correlation goes over to the value 2 for zero delay, in accord with the earlier experimental result obtained by Rebka and Pound (1957).

Concerning the description of optical coherence a new era started in 1963 when the path-breaking papers by Glauber (1963a-b) appeared on the quantum theory of optical coherence. He introduced the quantum coherence functions, which are in principle suitable to describe any interference phenomena or correlations of the electromagnetic radiation for an arbitrary spatio-temporal arrangement of the photodetectors. We mention that still nowadays there are details to be worked out on this subject, as is shown, for instance by the work of Schukin and Vogel (2006) appeared quite recently on the universal measurement of quantum correlation of radiation. Going back to the mid-sixties, we mention that many analyses appeared after the works of Glauber was published, e.g. the thoroughly written report by Mandel and Wolf (1965) on the coherence of optical fields and the work by Paul (1966) on the quantum theory of optical coherence. The later developments of this field can be kept track of e.g. in the books by Loudon (1973), Scully and Zubairy (1997) and Schleich (2001), which we have already quoted above, and by the references therein.

A fruitful impetus has been given to the investigations of correlation and quantum coherence of the optical field by the discovery of squeezed states, photon anti-bunching (see Kimble et al. 1977) and the observation of sub-poissonian photon statistics by Short and Mandel (1983). Such "non-classical effects" have been soon thoroughly analysed e.g. by Loudon (1980). Walls (1979) considered the new experimental results as "evidence for the quantum nature of light". In this context we also refer the reader to the excellent collection of papers published in the book edited by Dodonov and Man'ko (2003) on non-classical states of light, where one can find practically all the basic references. Nowadays all kinds of photon states can be produced at will (see e.g. Walther 2005a-b or Waks et al. 2006). Concerning the branch of single-photon sources and experiments we refer the reader to the focus issue in Volume 6 of the New Journal of Physics on "single photons on demand" (2004).

As is widely noticed in the physisist's community, the experiments on the wave-particle duality and, in particular, on entanglement (which we are not going to discuss in the present paper), deliver very counter-intuitive results. On the other hand we have a well-working formalism of the quantum theory of radiation, which always gives a recepie to calculate correctly the basic features of the experimental results. This, of course, does not mean that we really understand what is going on. There are still many different views concerning the photon concept. A nice collection of this views can be found e.g. in a recent special issue of OPN Trends **3** (2003) entitled "The Nature of Light. What is a Photon?" In particular, we refer the reader to the paper by Loudon (2003) and Muthukrishnan at al. (2003) appeared in this issue.

The motivation for writting up the present paper was to give a bit more intuitive description of the correlations appearing in "single-photon experiments". Our analysis is completely based on classical probability theory. We always assume that there is only one photon (or a wave packet of electromagnetic radiation) of energy $h\nu$ in the measuring apparatus during one elementary measurement. By fixing the the number of these elementary measurements we receive a strict anticorrelation of the the detector signals at the two arms of the measuring device because the photon (wave packet) cannot excite both of the detectors at the same time, due to the conservation of energy. This result is in complete accord with the experimental findings of Grangier et al. (1986). Then we consider the excitation of the measuring device (with two photodetectors placed on the opposite sides of a beam-splitter) by *serieses of n-photon sequences*. Now the joint probability distributions of the detection events can be represented as weighted sums (mixtures) of the joint distributions associated to the n-photon sequences of single-photon elementary measurements. It will turn out, that, though in each n-photon sequence strict anticorrelation exist between the detection events at the two photodetectors (which means that the normalized correlation coefficient is -1), negative, zero or positive correlation coefficients coming out for the serieses of n-photon sequences. We shall consider five cases of the excitation of the measuring device: the "number excitation", when it is secured that during one experimental run exactly n photons are falling into the measuring device. The next case is when we have a Poissonian sequence of such number excitations (this is the case of the "coherent light"). Then, we shall discuss the case of "thermal (or chaotic) excitation", when the weights are given by a Bose distribution of a black-body radiation. In this case we derive the photon-bunching, and the factor 2 for the normalized intensity-intensity correlation, which nowadays called the Hanbury Brown and Twiss effect. The next example will be the case of "squeezed excitation" when the weights of the n-photon sequences are given by the probability distribution of the photon numbers in a squeezed coherent state. The "photon anti-bunching" will be discussed, which has been set to be a genuine non-classical effect. Here we show that this effect can be obtained from classical probability theory. The last example refers to the "phase excitation", as we term. In this case the number of the single-photon experiments are distributed uniformly. We shall show that in this case the correlation coefficient is always positive (like in the case of the Bose excitation), but the normalised intensity-intensity correlation coefficient is smaller than 2 (Hanbury Brown and Twiss effect), namely, it is about 4/3.

**2. Examples of correlations in serieses of sequences of single-photon measurements**

Let us imagine the following experimental arrangement. The photon source sending off photons of energy $h\nu$ impinging on a beam-splitter which let them either passing through and detected by detector A, or they are reflected perpendicularly and detected by the detector B. The scheme of such an experiment is shown on Fig 1. Assume that the photons are coming such rarely, that during one single detection interval (*during one elementary experiment*) only one photon is present in the apparatus. Assume, moreover, that the photon energy $h\nu$ is only slightly larger than the excitation energy of both of the detectors. In this case, *due to the conservation of energy*, during one detection process either the detector A or the detector B can be excited (or none of them).

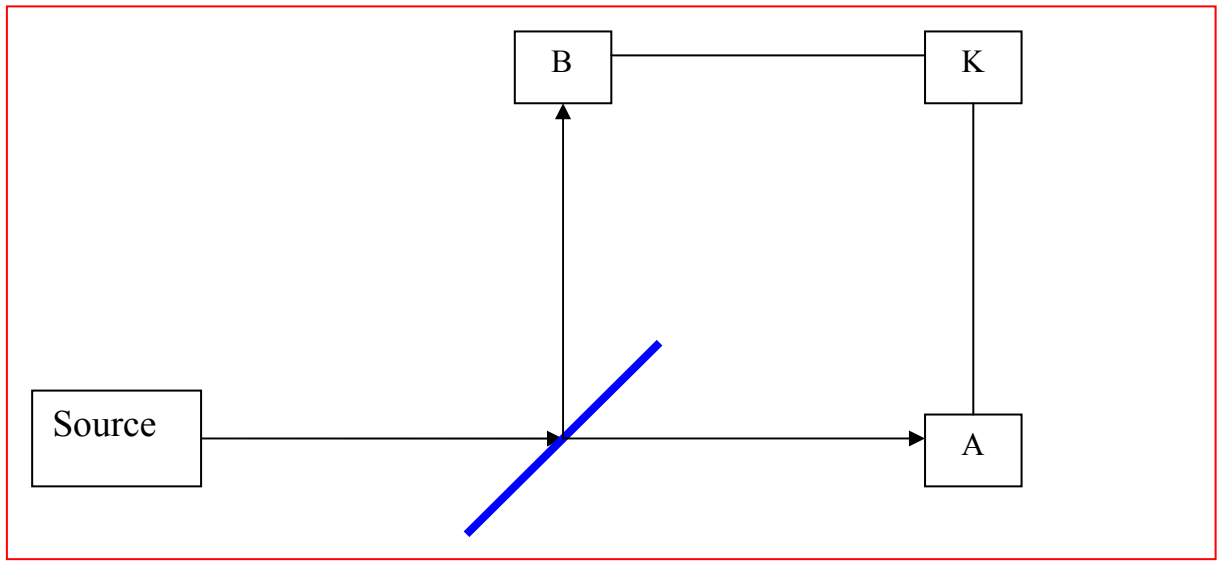

**Fig.1** Shows schematically the experimental arrangement under discussion. The source is assumed to send photons to the measuring apparatus such rarely, that there is always only one photon present during one elementary measurement. Since, according to our assumption, the energy of one photon is not enough to excite both of the detectors A and B, there is always a strict anticorrelation between the detections at A and B, during one elementary experiment.

Under this circumstances (because of the conservation of energy) it can never happen that both A and B are excited. The excitation of detectors A and B we call event *A* and *B*, respectively. The event when neither of the two detectors are excited will be called *C*. These three events are mutually exluding each other and the sum of them is the identity (absolutely certain) event *I*. The sum of the probabilities of this three events is clearly unity,

$$P(A) = p \ , \ P(B) = q \ , \ P(C) = r \ , \ p + q + r = 1. \tag{1}$$

In the following we shall assume that *r* is a non-zero quantity, i.e. there is always a certain finite probability that the photon is not detected in an elementary experiment. The probabilities *p*, *q* and *r* depend on the properties of the detectors and of the beam splitter. For instance, if A and B has the same properties (detection materials and efficiencies), and the beam splitter is of 50% transmittivity (50% reflectivity), then *q* must be equal to *p*.

To characterize the outcome of a *sequence* of *n* single-photon *independent elementary expriments* we introduce the random variables $\xi_n$ and $\eta_n$ by the following definitions. The variable $\xi_n(A)$ is the random number of independent elementary experiments (from altogether *n* experiments) in which detector A is excited. Similarly, the variable $\eta_n(B)$ is the number of independent elementary experiments (from altogether *n* experiments) in which detector B is excited. The joint distribution of these random variables is a trinomial distribution (Rényi 1962, p.118),

$$w_{mk}(n) \equiv P(\xi_n = m, \eta_n = k) = \frac{n!}{m!k!(n-m-k)!} p^m q^k r^{n-m-k}. \tag{2}$$

The above formula can be obtained by an elementary combinatorial calculation, where the order of the the results *A*, *B* and *C* within the sequence of the *n* independent elementary experiments is not taken into account.

In our following study we shall need the *normalized correlation coefficient R* of the detection results which is defined (Rényi 1962, p. 97) as

$$R(\xi_n, \eta_n) \equiv \frac{\overline{(\xi_n - \overline{\xi_n})(\eta_n - \overline{\eta_n})}}{\Delta \xi_n \Delta \eta_n} = \frac{\overline{\xi_n \cdot \eta_n} - \overline{\xi_n} \cdot \overline{\eta_n}}{\Delta \xi_n \Delta \eta_n}, \quad |R(\xi_n, \eta_n)| \leq 1,$$

$$(\Delta \xi_n)^2 \equiv \overline{(\xi_n - \overline{\xi_n})^2} \ , \quad (\Delta \eta_n)^2 \equiv \overline{(\eta_n - \overline{\eta_n})^2} \ , \tag{3}$$

where the upper dash denotes expectation value. In Eq. (3) we have introduced the dispersions $\Delta\xi_n$ and $\Delta\eta_n$ which are the positive square roots of the corresponding variances. In general, the calculations of expectation values and higher moments of probability distributions can be conveniently done by using the *generating functions* (Rényi 1962, p. 118) which we shall also use in the following. The two-variable generating function of the joint distribution given by Eq. (2) reads

$$G_n(x,y) \equiv \sum_{m=0}^{n} \sum_{k=0}^{n} w_{mk}(n) x^m y^k = (px + qy + r)^n , \tag{4}$$

where $x$ and $y$ are in general complex numbers satisfying the relations $|x|\leq 1$ and $|y|\leq 1$. For instance, the expectation values of the random variables $\xi_n$ and $\eta_n$ can be expressed in terms of the first order partial derivatives of the generating function,

$$\overline{\xi_n} \equiv \sum_{m=0}^{n} \sum_{k=0}^{n} w_{mk}(n) m = \left[\frac{\partial G_n(x,y)}{\partial x}\right]_{x=1, y=1} = np ,$$

$$\overline{\eta_n} \equiv \sum_{m=0}^{n} \sum_{k=0}^{n} w_{mk}(n) k = \left[\frac{\partial G_n(x,y)}{\partial y}\right]_{x=1, y=1} = nq , \tag{5}$$

and the higher moments can be calculated from the higher order derivatives. The variances of $\xi_n$ and $\eta_n$, and the expectation value of the product of them are given as

$$\Delta\xi_n^2 = np(1-p) \ , \quad \Delta\eta_n^2 = nq(1-q) \ ,$$

$$\overline{\xi_n \cdot \eta_n} = p\frac{\partial}{\partial p} q\frac{\partial}{\partial q}(p+q+r)^n = pqn(n-1) , \quad \frac{\overline{\xi_n \cdot \eta_n}}{\overline{\xi_n} \cdot \overline{\eta_n}} = 1 - \frac{1}{n} , \tag{6}$$

where we have also used the last equation in Eq. (1). The last equation of Eq. (6) shows that the ratio of the expectation value of the product of the random variables to the product of their expectation values is smaller then unity. Thus, according to Eq. (3), the correlation coefficient is negative. From Eqs. (5) and (6) we can express the normalized correlation coefficient of $\xi_n(A)$ and $\eta_n(B)$ defined in Eq. (3),

$$R^{number}(\xi_n, \eta_n) = -\sqrt{\frac{pq}{(1-p)(1-q)}} = -\frac{p}{1-p} . \tag{7}$$

The last equation of Eq. (7) is valid in the special case when $p = q$. The superscript "number" refers to that we are dealing here with a *fixed number of elementary experiments*. It is interesting to note that the normalized correlation coefficient does not depend on the total number of photons detected. The *negative correlation* or *anticorrelation* shown by Eq. (7) is in complete accord with our intuition and with the experimental results by Grangier et al. (1986). Since, during one elementary experiment only one photon is present in the measuring apparatus, then, if once this photon is absorbed by detector A (the event *A* results), owing to the conservation of energy, the detector B cannot be excited, so the event *B* can by no means result. Hence there is always a strict anticorrelation between $\xi_n(A)$ and $\eta_n(B)$ in a sequence of independent elementary experiments. Equation (7) quantitatively expresses this intuitively clear expectation. As we shall see in the following, the situation completely changes if we consider the outcome of *serieses* (*mixtures*) *of sequences* of the *n* elementary experiments with weights $W_n$.

By now we have assumed that the number of photons impinging into the apparatus is a *fixed number n*, so in the whole experimental run we can observe at most *n* detection event (this case has been experimentally realized by Grangier et al. (1986)). In the following we consider the case when we cannot secure this latter condition satisfied, but assume that the number of the elementary experiments (i.e. the number of photons) is a random variable with certain distributions characterized by the weights $\{W_n, n=0, 1, 2, …\}$, and we do not know in which sequence the photons were detected. In this case the possible outcomes of the whole experimental run can be represented by two random variables $\xi(A)$ and $\eta(B)$. The variable $\xi(A)$ is the random number of independent elementary experiments in which the detector A is excited. Similarly, the variable $\eta(B)$ is the random number of independent elementary experiments in which the detector B is excited. The joint distribution of $\xi$ and $\eta$ is given by the *mixture*

$$P(\xi = m, \eta = k) = \sum_{n=0}^{\infty} W_n P(\xi_n = m, \eta_n = k), \text{ with } \sum_{n=0}^{\infty} W_n = 1. \qquad (8)$$

According to Eq. (8), the generating function of the mixture is the weighted sum of the generating functions of the original distributions (Rényi 1962, p.117), i.e.

$$G(x, y) = \sum_{n=0}^{\infty} W_n G_n(x, y). \qquad (9)$$

As a first example, let us consider the case where the weights follow a Poisson distribution of parameter $\lambda$,

$$W_n^{Poisson} = \frac{\lambda^n e^{-\lambda}}{n!}. \qquad (10)$$

This case we term "Poisson excitation" of the measuring apparatus. The weights in Eq. (10) can be considered as the photon number distribution in a coherent electromagnetic radiation field (see Glauber 1963a-b), i.e., in a laser field. By using Eqs. (4), (9) and (10) we obtain (see Rényi 1962, p.119)

$$G^{Poisson}(x, y) = \exp[\lambda p(x-1)] \cdot \exp[\lambda q(y-1)], \qquad (11)$$

that is, the generating function is factorized into the product of two generating functions of Poisson distributions of parameters $\lambda p$ and $\lambda q$. This means that the random variables $\xi$ and $\eta$ are *independent* (not only *uncorrelated*) Poissonian variables, i.e. their correlation coefficient is zero,

$$\overline{\xi} = \lambda p, \quad \Delta\xi^2 = \lambda p, \quad \overline{\eta} = \lambda q, \quad \Delta\eta^2 = \lambda q, \quad \frac{\overline{\xi \cdot \eta}}{\overline{\xi} \cdot \overline{\eta}} = 1. \qquad (12)$$

From Eq.(12) we have

$$R^{Poisson}(\xi, \eta) = 0. \qquad (13)$$

Thus, the Poisson excitation makes the original negative correlation to increase up to zero. In the Ádám et al. (1955) experiment they measured a correlation less than 0.6%, so one could conclude that "the photon does not split". Brannen and Ferguson (1956) reinvestigated the same problem with the same conclusion (see also Farkas et al. 1964 and Arecchi et al. 1966). In fact, as we mentioned already in the introduction, Clauser (1974) has shown later, that the above experiment had not been conclusive for technical reasons, and he developed a more sophisticated experimental arrangement based on four photomultiplier. In our scheme we need not explicitely assume that the photon does not split, we merely assume that if it does not split, then it is *absorbed as a whole* ( according to Einstein, 1905) by one of the the detectors. On the other hand, if the packet of energy $h\nu$ (as a classical wave packet) does split at the beam-splitter, then neither of the detectors can be excited, because of the lack of the threshold energy which can excite them simultaneously. This is the case when event $C$ results, which can also happen when the photon does not split but it is simply not "observed" by the detectors for some other reasons.

Our next example is the "thermal excitation" characterized by the Bose distribution

$$W_n^{thermal} = (1-b)b^n = \frac{\overline{n}^n}{(1+\overline{n})^{1+n}}, \quad b \equiv \exp(-h\nu/kT), \quad \overline{n} = \frac{1}{e^{h\nu/kT}-1}, \qquad (14)$$

S. Varró: Correlations in single-photon experiments                                   8where $k$ denotes the Boltzmann constant and $T$ is the absolute temperature of a black-body radiation. According to Eqs. (4), (9) and (14), the generating function now reads

$$G^{thermal}(x,y) = \frac{1-b}{1-b(px+qy+r)}. \tag{15}$$

With the help of this generating function we derive the moments we need,

$$\overline{\xi} = \frac{pb}{(1-b)} = p\overline{n} \;,\; \Delta\xi^2 = \overline{\xi} + \overline{\xi}^2 \;,\; \overline{\eta} = \frac{qb}{(1-b)} = q\overline{n} \;,\; \Delta\eta^2 = \overline{\eta} + \overline{\eta}^2,$$

$$\overline{\xi\cdot\eta} = \frac{2pqb^2}{(1-b)^2} = 2pq\overline{n}^2 \;,\; \frac{\overline{\xi\cdot\eta}}{\overline{\xi}\cdot\overline{\eta}} = 2. \tag{16}$$

The normalized correlation coefficient can be calculated on the basis of Eq. (16), yielding

$$R^{thermal}(\xi,\eta) = \frac{pq}{\sqrt{p^2 + p(e^{h\nu/kT}-1)}\sqrt{q^2 + q(e^{h\nu/kT}-1)}} = \frac{p\overline{n}}{1+p\overline{n}}, \tag{17}$$

where we have taken into account the definition of the parameter $b$ given in Eq. (14). The last equation of Eq. (17) is valid in the special case when $p = q$. The second and the fourth equations of Eq. (16) show that in the variances of the counts both the "particle-like fluctuation term" and the "wave-like fluctuation term" are present, like in Einstein's famous fluctuation formula (Einstein 1909a). The physical content of these terms in the case of black-body radiation has recently been discussed in details in Varró (2006). The last equation in Eq. (16) expresses the well-known "photon bunching" or Handbury Brown and Twiss (1956) effect, which was also measured by Rebka and Pound (1957) and Arecchi et al. (1966). It is remarkable that according to our original assumption, there is always only one single photon (one single energy element $h\nu$ available) in the measuring apparatus, hence there is a strict anticorrelation between the detection events during a sequence of elementary experiments with a fixed photon number, as is shown by Eq. (7). In spite of this assumed spatial separation, in a random series of such sequences distributed according to Eq. (14), the randomness of the excitation produces a positive correlation of the counts, as we see in Eq. (17).

Our next example is the "squeezed excitation" of the apparatus, where the weights have the form

$$W_n^{squeezed} = \sqrt{1-\varsigma^2}\exp\left[-a^2(1+\varsigma)\right]\frac{\varsigma^n}{2^n n!}H_n^2\left(a\frac{1+\varsigma}{\sqrt{2\varsigma}}\right) \;,\; 0<a \;,\; 0<\varsigma<1, \tag{18}$$

where $H_n$ denotes the $n$-th Hermite polynomial. The probability distribution given by Eq. (18) is the photon number distribution of a squeezed coherent state (see e. g. the thoroughly written recent review by Wünsche 2003), where $a$ is the displacement parameter and $\zeta$ is the squeezing parameter. For later convenience we introduce an alternative parametrization for the measure of the squeezing of the excitation used by Schleich and Wheeler (1987a-b) which suits better in considering the case of high squeezing,

$$\varsigma \equiv \frac{s-1}{s+1} \;,\; 1<s \;,\; \varepsilon \equiv \frac{2}{s} \;,\; \varepsilon<2. \tag{19}$$

The case of high squeezing corresponds to $s \gg 1$ and $\varepsilon \ll 1$. In order to calculate the generating function of the mixed joint distribution Eq. (9) for the present case we use Mehler's formula (Erdélyi 1953)

$$\sum_{n=0}^{\infty}\frac{t^n}{2^n n!}H_n(\alpha)H_n(\beta) = \frac{1}{\sqrt{1-t^2}}\exp\left[\frac{2\alpha\beta t - (\alpha^2+\beta^2)t^2}{1-t^2}\right] = \frac{1}{\sqrt{1-t^2}}\exp\left(2\alpha^2\frac{t}{1+t}\right), \tag{20}$$

where the last equation holds for $\alpha=\beta$. With the help of Eqs. (4), (9), (18) and (20) we have

$$G^{squeezed}(x,y) = \sqrt{1-\varsigma^2} \exp\left[-a^2(1+\varsigma)\right] \frac{1}{\sqrt{1-t^2}} \exp\left(u \frac{t}{1+t}\right), \text{ where}$$

$$t \equiv \varsigma(px+qy+r), \quad u \equiv a^2(1+\varsigma)^2/\varsigma. \tag{21}$$

By using the generating function Eq. (21), the moments of the random variables $\xi$ and $\eta$ can be obtained by simple differentiations,

$$\overline{\xi} = p\left(a^2 + \frac{\varsigma^2}{1-\varsigma^2}\right), \quad \overline{\eta} = q\left(a^2 + \frac{\varsigma^2}{1-\varsigma^2}\right), \tag{22}$$

$$\Delta\xi^2 = p\alpha + p^2\beta, \quad \alpha \equiv a^2 + \frac{\varsigma^2}{1-\varsigma^2}, \quad \beta \equiv \left[\varsigma^2 \frac{1+\varsigma^2}{(1-\varsigma^2)^2} - 2\varsigma \frac{a^2}{(1+\varsigma)}\right], \tag{23}$$

and a similar expression holds for $\Delta\eta^2$ with $p$ replaced by $q$. The nominator of the normalized correlation coefficient reads

$$\overline{\xi \cdot \eta} - \overline{\xi} \cdot \overline{\eta} = pq\beta, \tag{24}$$

which, according to the definition of $\beta$, can either be positive or negative (or zero) depending on the values of the displacement parameter $a$ and the squeezing parameter $\varsigma$. Thus the "squeezed excitation", Eq. (18), can cause both "bunching" and "anti-bunching" of the counts in a random series of sequences of elementary single-photon experiments (concerning the first experimental results on anti-bunching and sub-poissonian photon statistics see Kimble et al. 1977 and Short and Mandel 1983). From Eqs. (23) and (24) the normalized correlation coefficient can be obtained,

$$R^{squeezed}(\xi,\eta) = \frac{pq\beta}{\sqrt{p\alpha+p^2\beta}\sqrt{q\alpha+q^2\beta}} = \frac{p\beta}{\alpha+p\beta}, \tag{25}$$

where the last equation is valid for $p = q$. In the case of high squeezing, when $\varsigma$ approaches unity ($1 \ll s$, i.e. $\varepsilon \ll 1$), the subsidiary parameters $\alpha$ and $\beta$, defined in Eq. (23), can be well approximated as $\alpha \approx (1/2\varepsilon) + a^2$ and $\beta \approx (1/\varepsilon^2) - a^2$, hence the correlation coefficient becomes

$$R^{squeezed}(\xi,\eta) \approx \frac{p(1-\varepsilon^2 a^2)}{\varepsilon/2 + \varepsilon^2 a^2 + p(1-\varepsilon^2 a^2)} \approx \frac{p(1-\varepsilon^2 a^2)}{\varepsilon^2 a^2 + p(1-\varepsilon^2 a^2)}, \quad (1 \ll s, \varepsilon \ll 1). \tag{26}$$

The last relation has been obtained by assuming, in addition, that $1 \ll 2\varepsilon a^2$. We note that if the latter condition is satisfied then the average number of counts approximately equals $pa^2$. The denominator in Eq. (26) is always positive, so the sign of the correlation coefficient is determined by the magnitude of $\varepsilon^2 a^2$. For $1 \ll \varepsilon^2 a^2$ we get back the maximum anticorrelation expressed by Eq. (7). If $\beta$ is zero, then, according to Eq. (25), the number of counts at detectors A and B are not correlated, like in the case of Poisson excitation (see Eq. (13)). However, this does not mean that $\xi$ and $\eta$ are independent, since the generating function, Eq. (21), is now clearly not factorized as there. For highly squeezed excitations Eq. (26) expresses "bunching" when $\varepsilon a < 1$, and it expresses "anti-bunching" when $\varepsilon a > 1$.

In our last example we consider the finite uniform distibution of the sequences of elementary experiments,

$$W_n^{phase} = \frac{1}{N+1}, \quad (n=0,1,2...,N). \tag{27}$$

The superscript "phase" in Eq. (27) refers to the (uniform) photon number distribution in phase states introduced by Loudon (1973). The generating function of the joint distribution of $\xi$ and $\eta$ now reads

$$G^{phase}(x,y) = \frac{1}{N+1} \sum_{n=0}^{N} G_n(x,y) = \frac{1}{N+1} \cdot \frac{1-(px+qy+r)^{N+1}}{1-(px+qy+r)}, \tag{28}$$

where we have used Eq. (4). On the basis of Eq. (28) the moments of the distribution can be easily calculated,

$$\bar{\xi} = pN/2 \ , \ \Delta\xi^2 = [pN(6-4p) + p^2N^2]/12 \ , \ \bar{\eta} = qN/2 \ , \ \Delta\eta^2 = [qN(6-4p) + q^2N^2]/12,$$

$$\overline{\xi \cdot \eta} = \frac{N}{2}\frac{4N-1}{6}pq \ , \ \frac{\overline{\xi \cdot \eta}}{\bar{\xi}\cdot\bar{\eta}} = \frac{4}{3} - \frac{1}{3N} \ . \tag{29}$$

The last equation in Eq. (29) shows that in the case of "phase excitation", Eq. (27), the counts in detector A and B are always positively correlated. However the degree of this correlation ($\approx 4/3$) can never reach the value 2 (see the last equation in Eq. (16)) of the classic bunching (Hanbury-Brown and Twiss effect), appearing in the case of thermal excitation. From Eq. (29) the normalized correlation coefficient is obtained,

$$R^{phase}(\xi,\eta) = \frac{pq(N^2-N)}{\sqrt{p(6-4p)N+p^2N^2}\sqrt{q(6-4q)N+q^2N^2}} = \frac{p(N^2-N)}{(6-4p)N+pN^2} \ , \tag{30}$$

where the last equation is valid in the case when $p = q$. In this special case, of course, $p < ½$, because our original constraint condition in Eq. (1) has to be satisfied: $(2p+r)=1$ with $0 < r$. For very large excitations ($N \gg 1$) the correlation coefficient approaches +1.

To conclude the present paper, let us summarize in Table 1 the results on the correlations, obtained in the five examples considered above.

| number | Poisson | thermal | squeezed | phase |
|---|---|---|---|---|
| $pn$ | $p\lambda$ | $p\bar{n}$ | $p(a^2 + 1/2\varepsilon)$ | $pN/2$ |
| $1 - \frac{1}{n}$ | 1 | 2 | $1 + 4\frac{1-\varepsilon^2 a^2}{(1+2\varepsilon \cdot a^2)^2}$ | $\frac{4}{3} - \frac{1}{3N} \approx \frac{4}{3}$ |
| $-\frac{p}{1-p}$ | 0 | $\frac{p\bar{n}}{1+p\bar{n}}$ | $\frac{p(1-\varepsilon^2 a^2)}{\varepsilon/2 + \varepsilon^2 a^2 + p(1-\varepsilon^2 a^2)}$ | $\frac{p(N^2-N)}{(6-4p)N+pN^2}$ |

**Table 1.** In the second raw the average number of counts $\bar{\xi}$ at the detector A are shown on the basis of Eqs. (5), (12), (16), (22) and (29). In the third raw the normalized intensity-intensity correlation $(\overline{\xi \cdot \eta})/(\bar{\xi}\cdot\bar{\eta})$ is displayed for the different types of excitations, according to Eqs. (6), (12), (16), (24) and (29). In the fourth raw the corresponding normalized correlation coefficients $R(\xi,\eta)$ are listed on the basis of Eqs. (7), (13), (17), (26) and (30) in the special case when $p = q$. For the sake of simplicity, in the case of "squeezed excitation" we presented the approximate formulae valid for large squeezing ($1 \ll s = 2/\varepsilon$).

### 3. Summary

As is clearly seen in Table 1., the strict anticorrelation (see also the experimental results by Grangier et al. 1986) obtained for a fixed number of elementary single-photon experiments (first column) is considerably modified if we evaluate the counts of the detectors for random serieses of experiments. In the latter case the length $n$ of a sequence of measurements is a random variable whose distribution is governed by the characteristics of the source. Though, in this case too, during one elementary measurement still only one photon is present in the apparatus, the *measured* correlation coefficient $R(\xi,\eta)$ can take on either positive or negative values, or it can be exactly zero, like in the case of Poisson excitation (second column). In the latter case, regardless of the magnitude of the intensity of the excitation $\lambda$, the number of counts at the detectors A and B are independent random variables. This result is in agreement with the experimental findings by Ádám et al. (1955), Arecchi et al. (1966) and Clauser (1974). The number of counts $\xi$ and $\eta$ can be uncorrelated in the case of "squeezed excitations" (fourth column) if $\varepsilon a = 1$, but this does not mean, that these random variables are independent, since the generating function, Eq. (21), does not factorize (which is a necessary but not sufficient condition for independence). Here the counts can also show bunching ($\varepsilon a < 1$) or anti-bunching ($\varepsilon a > 1$), depending on the parameters of the excitation of the measuring apparatus. For the "thermal or Gaussian excitations" and for the "phase excitations" (the third and the fifth columns, respectively) the correlations are always positive. Regardless of the size of the average counts (intensities), the normalized intensity-intensity correlations are 2 (Hanbury Brown and Twiss effect 1956, see also Arecchi et al. 1966) and 4/3, respectively.

**Acknowledgements.** This work has been supported by the Hungarian National Scientific Research Foundation (OTKA), Grant No. T048324.